\documentclass[twocolumn]{svjour3}          
\usepackage{graphicx}
\usepackage[numbers]{natbib}
\usepackage{amsmath,amssymb}

\journalname{Granular Matter}

\begin{document}

\title{Friction and Pressure-Dependence of Force Chain Communities in Granular Materials \thanks{This research has been supported by the James S. McDonnell Foundation and the National Science Foundation (DMR-1206808).}}


\author{Yuming Huang \and Karen E. Daniels }


\institute{Yuming Huang \at
              Dept. of Physics, North Carolina State University \\
              Raleigh, NC, USA
           \and
           Karen E. Daniels \at
              Dept. of Physics, North Carolina State University \\
              Raleigh, NC, USA \\
              \email{kdaniel@ncsu.edu} 
}

\date{Received: \today / Accepted: date} 

\maketitle

\begin{abstract}
Granular materials transmit stress via a network of force chains. Despite the importance of these chains to characterizing the stress state and dynamics of the system, there is no common framework for quantifying their their properties. Recently, attention has turned to the tools of network science as a promising route to such a description. 
In this paper, we apply community detection techniques to numerically-generated  packings of spheres over a range of interparticle friction coefficients and confining pressures. In order to extract chain-like features, we use a modularity maximization with a recently-developed geographical null model \cite{Bassett2015}, and optimize the technique to detect branched structures by minimizing the normalized convex hull of the detected communities. 
We characterize the force chain communities by their size (number of particles), network strength (internal forces), and normalized convex hull ratio (sparseness). We find the that the first two exhibit an approximately linear correlation and are therefore largely redundant. For both pressure $P$ and interparticle friction $\mu$, we observe crossovers in behavior. For $\mu \lesssim  0.1$, the packings exhibit more sensitivity to pressure. In addition, we identify a crossover pressure where the frictional dependence switches from having more large/strong communities at low $\mu$ vs. high $\mu$. We explain these phenomena by comparison to the spatial distribution of communities along the vertical axis of the system. These results provide new tools for considering the mesoscale structure of a granular system and pave the way for reduced descriptions based on the force chain structure.

\end{abstract}

\section{Introduction}
\label{intro}

For more than half a century, it has been common to visualize the heterogeneous force transmission in granular materials  \cite{Dantu1957,Liu1995,Howell1999}: these patterns have come to be known as {\it force chains.} It has been less clear, however, how best to provide a mesoscale description of this network of interparticle contacts. A better understanding of the important length scales over which intermediate structures are present would provide new routes to connect particle-scale properties to bulk properties. In this paper, we take the growing field of network science as our inspiration \cite{NewmanBook}. Network techniques can be applied to such varied systems as social networks, neural systems, or airline route maps: anything which can be reduced to a network of nodes and the edges (links) that connect them. In the sphere packings studied here, the network is composed of a set of particles (nodes) and the normal contact force between adjacent particles (edges).

The use of network science techniques has attracted significant attention in the past few years, particularly as a way to extract the ``backbone'' of the most important particles in the system, and to follow the evolution of those networks under loading. By considering the entire network \cite{Bassett2015,Arevalo2010,Walker2010,Walker2012,Bassett2012} it is possible extract statistics about the degree of connectivity among the particles, and how that influences the bulk response. 

One approach has been to define the main network based on  a set of rules about how particles in a force chain should be connected. For example, \citet{Peters2005a} and \citet{Zhang2013} define the backbone by setting a threshold value for contact forces and the angles between particles, while  \citet{Kondic2012} use a topological invariant called the zeroth Betti number \cite{kaczynski2004computational} to characterize the size of connected clusters. A disadvantage of these approaches is that  thresholding strictly removes the weak interparticle contacts from consideration, even though of these forces are thought to play an important role in providing lateral stability \cite{Radjai1998a}. To address this issue, it is possible to use community-detection techniques \cite{Porter2009,Fortunato2010} that allow for optimized partitioning into clusters without a hard threshold. For example, \citet{Navakas2010,Navakas2014} and \citet{Bassett2012} identified force clusters which have stronger interparticle forces within 
each cluster 
than 
between them. However, the force chain communities detected in this way take the form of compact domains rather than branched networks. Therefore, they seem unlikely to be the correct mesoscale units from which to explain sound propagation \cite{Owens2011} (observed to be along force chains) or the changing network of contacts under flow \cite{Herrera2011} (they form a giant component of broken links). 

Recently, \citet{Bassett2015} recognized that community-detection algorithms which depend on a random null model miss an important aspect of granular materials: that grains are geographically constrained to be connected only to their neighbors. By working with a null model that respects these geographic constraints, the detected communities  take the form of the expected branched structures. This {\it geographical null model}, when applied to either simulated (frictionless)  or laboratory (frictional) packings of disks, was able to distinguish the different force chain morphologies of the two distinct datasets.

In this paper, we adapt this technique to perform a similar community detection in simulations of 3D granular packings, and examine how these communities systematically change as a function of confining pressure and interparticle friction. 
By using simulations, we can test the methods in a controlled environment where it is possible to generate many independent realizations. Increasingly, such data is becoming available in 3D granular experiments \cite{Mukhopadhyay2011,Saadatfar2012,Brodu2015}, in the form of normal contact forces measured from the macroscopic deformations of soft particles.

The community detection method consists of two steps, modularity maximization to partition the network into clusters, and selecting a resolution parameter which controls the total number and shape of these clusters. To perform the maximization, we use the same geographic null model as in \citet{Bassett2015}, allowing us to incorporate contact information. In selecting a resolution parameter, we found that the previous technique \cite{Bassett2015} was inadequate for 3D systems. Therefore, we developed a new figure of merit to quantify the degree of branching within the communities, based on the convex hull of the constituent particles.

When the process is complete, each packing is partitioned into a set of branched communities (with many interstitial communities consisting of  only a few weak-force particles). We characterize the ensemble of communities by their size, strength, and degree of branching as a function of both interparticle friction and pressure. Both friction and pressure influence the network properties of force chains, and we observe crossovers as a function of both parameters.

\begin{figure}
\centerline{\includegraphics[width=\linewidth]{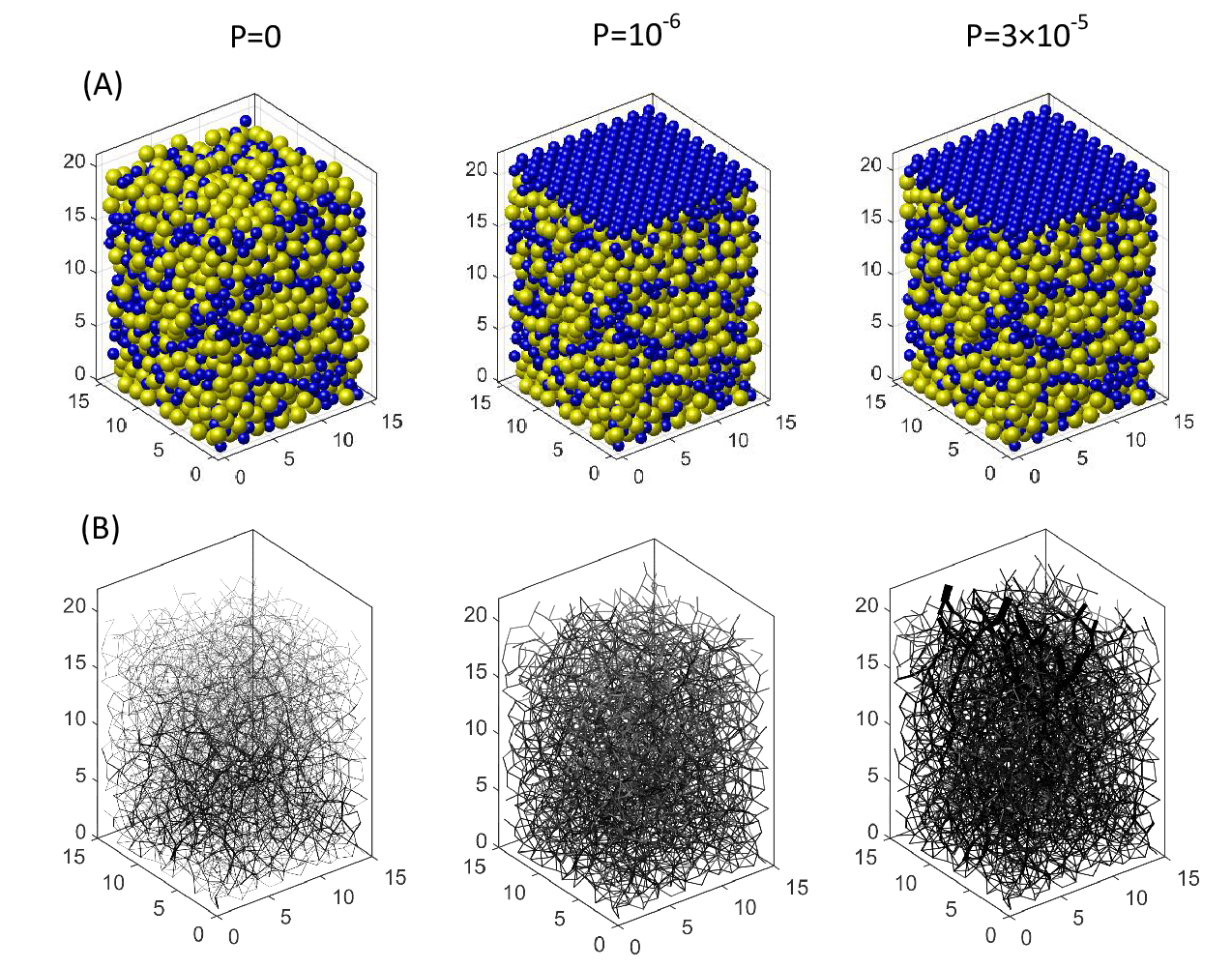}}
\caption{(A) Sample particle configurations generated using LAMMPS, with  small particles (diameter $d$)  shown in blue and large particles ($1.4d$) shown in yellow. In subsequent steps, we apply pressure from above using a flat slab. 
(B) Corresponding normal force networks, with bar thickness proportional to the normal force magnitude.}
\label{fig:lammps}
\end{figure}

\section{Simulation Methods \label{Simulation}}

We perform our numerical simulations using the discrete element model LAMMPS (Large-scale Atomic/ Molecular Massively Parallel Simulator) \cite{lammps} maintained by Sandia National Labs.  This open-source software is based on a fast parallel algorithm \cite{Plimpton1995} for molecular dynamics. Our simulations contain $N = 3000$ bidisperse spheres poured from above, half of diameter $d$ and half of diameter $1.4d$. The simulation cell has lateral dimensions $15d \times 15d$ (periodic boundary conditions in both directions) and height $26d$ (open at the top, closed at the bottom) with gravity acting downwards. To model a hard, frictional granular material, we use a Hertzian contact model with a normal elastic constant $K_n = 2 \times 10^5$ and a tangential elastic constant $K_t = \frac{2}{7} K_n$. These values, when re-dimensionalized, approximately correspond to ruby spheres  \cite{Chen2006} of centimetric size. We vary the interparticle friction coefficient over seven values $\mu =  0, 0.01, 0.03, 0.1, 0.
3, 1, 3$ to examine the dependence of our results on interparticle friction.

We mimic  experimental protocols in which particles are  poured  into a box from above, and then compressed via a uniform pressure. Our numerical pouring method, adapted from \citet{Silbert2002a}, mimics pouring particles through a sieve (to prevent the formation of a conical heap) by generating particles at random positions at the top of the simulation volume and allowing them to fall downward under the force of gravity. We select a very low packing fraction $\phi_i = 0.005$ for this insertion region, so that the resulting mean coordination number ${\bar c} = 5.5$ is independent of the choice of $\phi_i$ \cite{Blumenfeld2005}. After inserting all particles into the container, we allow the kinetic energy to dissipate until it is less than $10^{-8}\, mgd$ \cite{Silbert2002a}.

For each such initial packing, we apply pressure by generating a massless slab (size $15d \times 15d \times 2d$, FCC lattice, as shown in Fig.~\ref{fig:lammps}) at the top of the packing. From this initial state (compressed by gravity, but no additional pressure) we apply an increasing series pressures ($P = 10^{-6}, 3\times10^{-6}, 10^{-5}, 3\times10^{-5}, 10^{-4}$, measured in units of $K_n$) to the upper surface of the resulting packing. The magnitude of the non-dimensionalized confining pressure is just below those reported in recent experiments on softer particles \cite{Owens2013}. After the kinetic energy of the system has again dissipated, we record position and force measurements before iterating through all 6 pressure values. At each step, we record all particle positions and interparticle forces; sample normal force networks are shown in Fig.~\ref{fig:lammps}.

For each of the seven values of $\mu$, we perform 20 independent simulations starting from different random initial conditions. We use a bootstrap-like process (sampling with replacement) to confirm that this is sufficient for reliable statistics. In a few places, noted within the text, the fluctuations were large enough that this criterion was not satisfied.
In all of our analyses, we consider only the normal component of the interparticle forces, as is currently measured in experiments \cite{Mukhopadhyay2011, Saadatfar2012,Brodu2015}. This simplification also allows us to directly compare both frictionless and frictional packings.

\section{Community Detection \label{CD}}

Our goal is to partition the granular packing into {\it communities} of particles which have high interparticle forces internally (locally stiff) and low interparticle forces in their connections to other communities. We build on the work of \citet{Bassett2015,Bassett2012}, which utilizes the open source network analysis tool {\tt GenLouvain} (Version 2.0) from NetWiki \cite{InderjitS.Jutla} to implement the modularity maximization method \cite{Newman2004a} of community detection.

The network analysis begins from a representation of the normal force network (Fig.~\ref{fig:lammps}b) as an $N\times N$ weighted adjacency matrix $\bf W$. Each element $W_{ij}$ is zero for particles not in contact, and  $f_{ij}/\bar{f}$ for all non-zero interparticle forces (scaled by the mean normal force $\bar{f}$ for the whole packing). The modularity $Q$ of a network is a scalar value calculated from 
\begin{equation}
Q=\sum_{i,j}  \left[ W_{ij}-\gamma P_{ij} \right] \delta (c_{i},c_{j}) 
\label{eq:mod}
\end{equation}
where $\gamma$ is a resolution parameter, $P_{ij}$ is the expected weight of an edge due to a specific null model, $c_i$ and $c_j$ are the (numbered) community assignments for particle $i$ and $j$, and $\delta$ is the  Kronecker delta function. If particles $i,j$ are assigned to in the same community, then  $\delta(c_{i},c_{j})=1$, otherwise $\delta(c_{i},c_{j})=0$. The optimization process adjusts the community assignments for fixed $\gamma$ and fixed null model. As developed in \citet{Bassett2015}, we utilize a physically-motivated {\it geographic null model} in which particles connect to a community through their direct neighbors: 
\begin{equation}
P_{ij}=\left\{\begin{matrix}
1,\qquad W_{ij}\neq 0,\\ 
0,\qquad W_{ij}=0.
\end{matrix}\right.
\label{eq:nullmodel}
\end{equation}
We chose not to use the more common Newman-Girvan null model \cite{Newman2004a} which allows for arbitrary connections between particles \cite{Bassett2012, Navakas2010} because in 2D packings of disks, the geographic null model has been shown to successfully generate communities with chain-like morphologies \cite{Bassett2015} (rather than compact domains).

Because modularity maximization (finding the largest value of $Q$) is an NP-hard problem, the published methods \cite{InderjitS.Jutla} use a greedy heuristic algorithm. To test the stability of this method, we run this algorithm 100 times on the same force network and find that the fluctuation of maximal value of the modularity $Q$ is within 1\%. In addition, we observe that the 15 largest communities consist of the same core group of particles: 70\% of the same subset of particles are included in 90\% of the iterations.  We additionally find that fluctuations in $Q$ are not accompanied by fluctuations in the morphology of the detected communities (to be quantified in \S\ref{sec:morph}). 

\begin{figure}
\centerline{\includegraphics[width=0.7\linewidth]{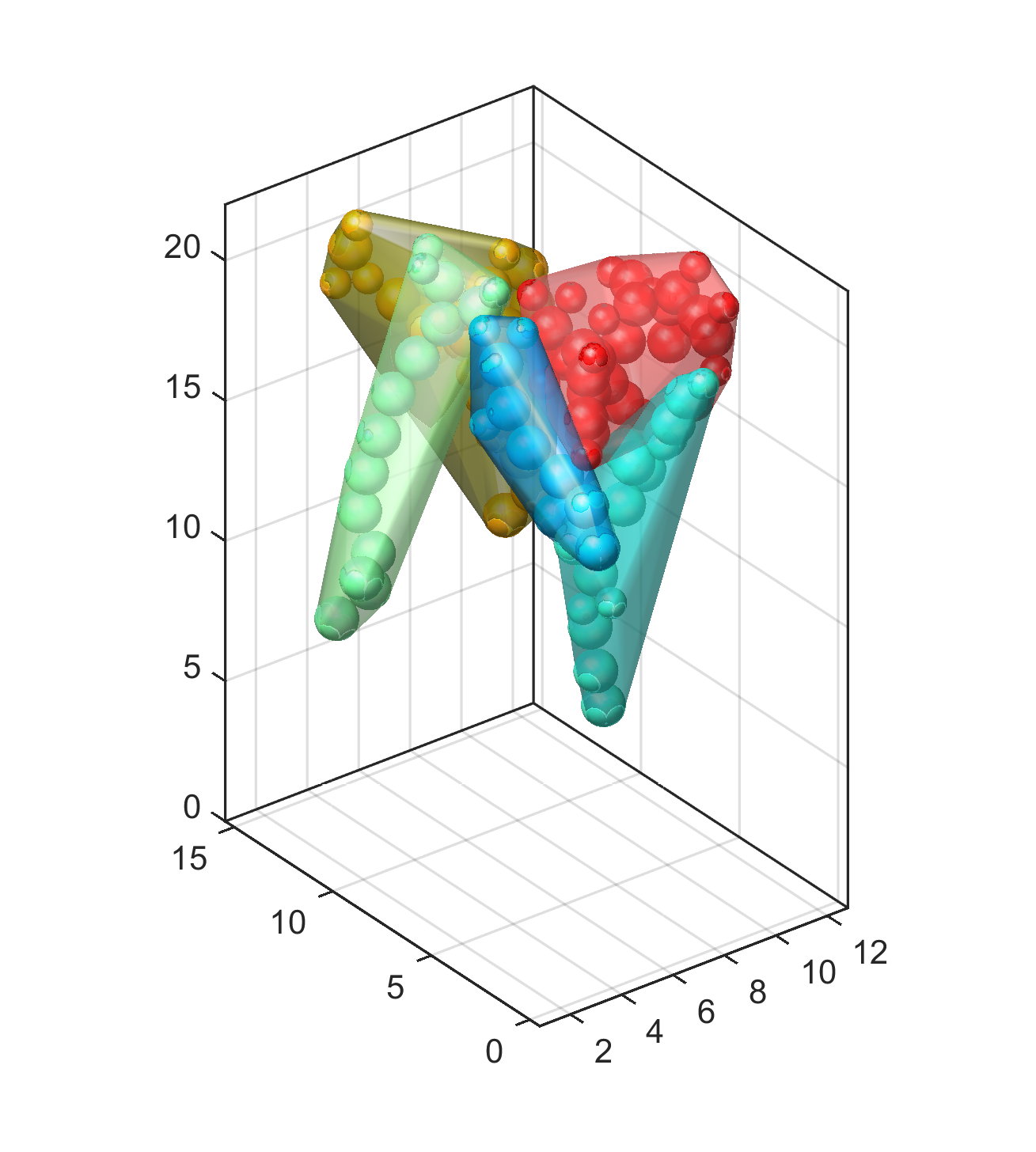}}
\caption{Five example communities ($\gamma = 3$) and their convex hulls, for a packing at with $\mu=0.3$ and  $P=10^{-4}$. Only communities containing more than 10 particles are shown.}
\label{fig:convexhull}
\end{figure}

The choice of resolution parameter $\gamma$ controls the total number of communities identified, and also their morphology. For  $\gamma < 1$, optimizing $Q$ favors large communities, while for $\gamma > 1$ small communities dominate. To select the optimal value of $\gamma$, we seek a figure of merit which quantifies the extent to which the detected communities take on a chain-like character: branched and sparse. We found that the technique used by \citet{Bassett2015} for 2D packings was ineffective in 3D systems. We therefore define a new figure of merit, the normalized convex hull ratio $H_c$
\begin{equation}
H_c = \frac{V_\mathrm{p}}{V_\mathrm{hull}} 
\label{eq:hr}
\end{equation} 
where $V_\mathrm{p}$ is the total volume of particles in the community and $V_\mathrm{hull}$ is the volume of the convex hull of the community. Branched/sparse communities will have lower values of $H_c$. 
To calculate $V_\mathrm{hull}$, we discretize each sphere as a $7\times7$ matrix of points and determine the convex hull using the Matlab {\tt boundary} function.  Fig.~\ref{fig:convexhull} shows example convex hulls.

\begin{figure}
\centerline{\includegraphics[width=\linewidth]{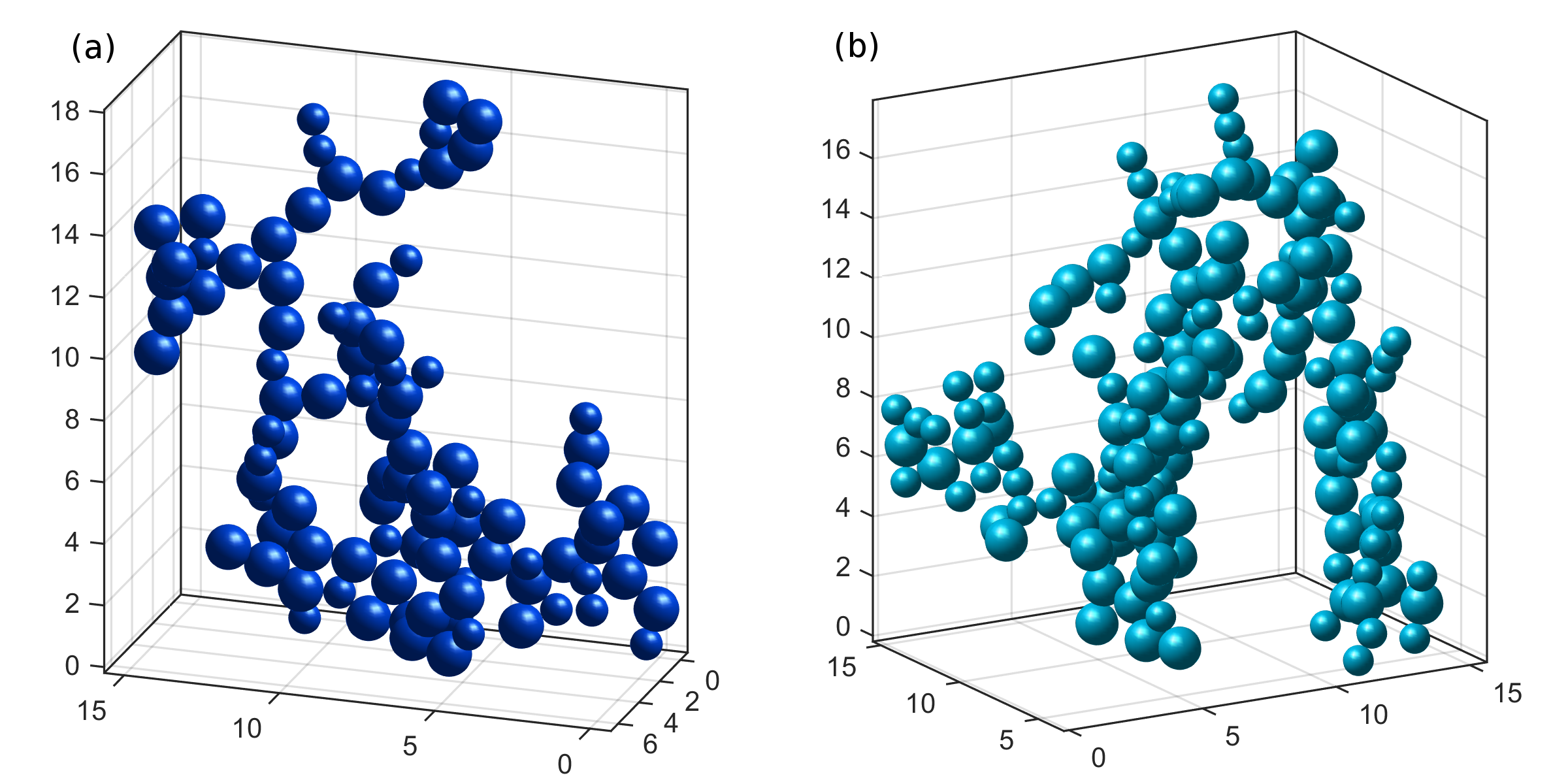}}
\caption{Examples of two medium size communities. Resolution parameter $\gamma=1.1$. $\mu=0.3$. $P=10^{-4}$.}
\label{fig:mid}
\end{figure}

Fig.~\ref{fig:mid} shows two examples of intermediate-size communities, shown in isolation to make them more visible. Note the chain-like structures dominating the communities, providing a sparse structure with a low hull ratio. The interstices of such communities can be filled either by smaller communities, or by intercollated communities which are also branched.

\begin{figure*}
\centerline{\includegraphics[width=0.7\linewidth]{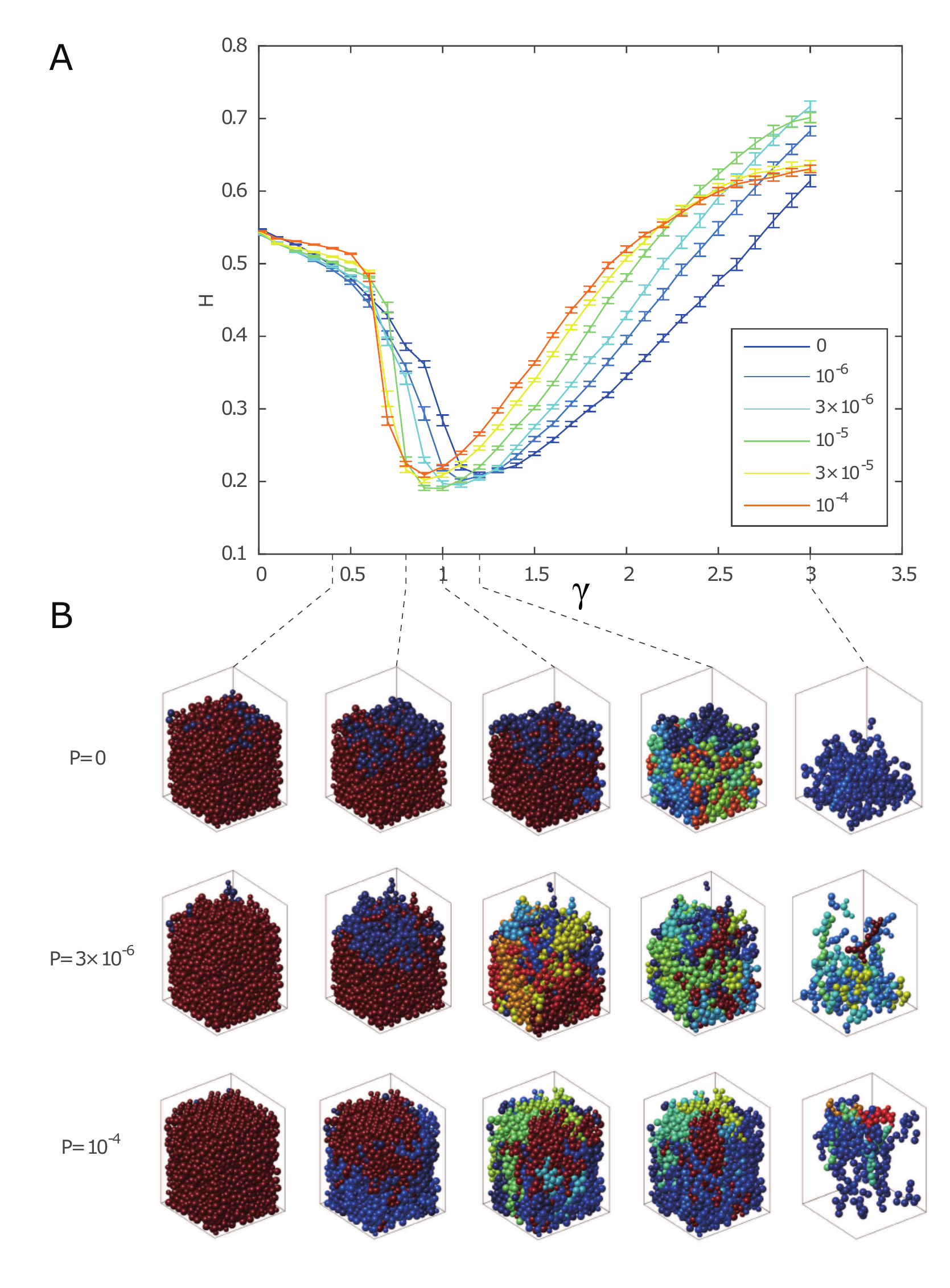}}
\caption{(a) Average hull ratio $H$ as a function of resolution parameter $\gamma$ for various pressures, with $\mu$=0.3. (b) Community detection results from selected pressures and $\gamma$ (0.4, 0.8, 1, 1.2, 3). The color scale for each packing ranges from zero (deep blue) to the its maximum value of  $\sigma_{c}$ (red); many blue particles are of similar color but do not below to the same community.  For clarity, communities with only 1 particle are hidden.}
\label{fig:optgamma}
\end{figure*}

For each packing, we calculate the mean hull ratio $H$ by averaging the measured $H_{c}$ weighted by the number of particles in each community, excluding communities which contain only one particle. To determine the optimal value of $\gamma$ to use in our analysis, we measure how $H$ changes as a function of $\gamma$ across a range of pressures.  As shown in Fig.~\ref{fig:optgamma}a, there is a clear minimum value of $H$ which is approximately consistent across different values of $P$. In the analysis below, we utilize $\gamma=1.1$ in all cases.

The effect of changing $\gamma$ can be understood by examining the contribution each community makes to the modularity $Q$. The network force
\begin{equation}
\sigma_{c} = \sum_{i,j\in C}[W_{ij}-\gamma P_{ij}] 
\label{eq:nf}
\end{equation} 
is the contribution to $Q$ (Eq.~\ref{eq:mod}) from only the particles located in a particular community $C$. Its value increases due to both the normal forces in the community being large, and from the size (number of nodes) $S_c$ in the community.

In Fig.~\ref{fig:optgamma}b, the individual communities are colored by their particular network force $\sigma_c$. For small $\gamma$, the $W_{ij}$ term dominates the sum in Eq.~\ref{eq:mod}, If $\gamma$ is small enough that $W_{ij}-\gamma P_{ij}$ is always positive, then the largest value of $Q$ is obtained by putting many particles in the same community ($\delta (c_{i},c_{j})$ is nearly always 1).
For larger $\gamma$, the null model $P_{ij}$ will have more influence on the chosen communities, and the particular geometry and interparticle forces matter. If $\gamma$ is large enough that $W_{ij}-\gamma P_{ij}$ is always negative, then the optimal $Q$ is zero by letting all $\delta (c_{i},c_{j})$ be zero. In that case, the optimum value of $Q$ is obtained when each community contains only a single particle. 

For the  special case where $\gamma=1$, contact forces between particles ($W_{ij}$) are directly compared with the average contact force ($P_{ij}$) in the system. The modularity $Q$ is increased when more particles with multiple contact forces greater than $\bar{f}$ are included in communities (force chains in our sense). The  choice of $\gamma=1$ is similar to finding force chains by thresholding at a minimum force, often set to be $\bar{f}$. However, in contrast with thresholding methods, the modularity maximization method is flexible rather than binary.

\section{Results \label{Results}}

Using these community detection methods, we describe how the force chain network changes as a result of both interparticle friction $\mu$ and confining pressure $P$.  For each community, we consider three properties: the community size $S_c$, the network force $\sigma_c$ (community strength), and the hull ratio $H_c$ (community morphology). In all cases, community-detection is performed at fixed resolution parameter $\gamma = 1.1$, chosen as a compromise value for the whole parameter regime. 

\subsection{Community Size and Strength} 

\begin{figure}
\centering
\includegraphics[width=0.4\textwidth]{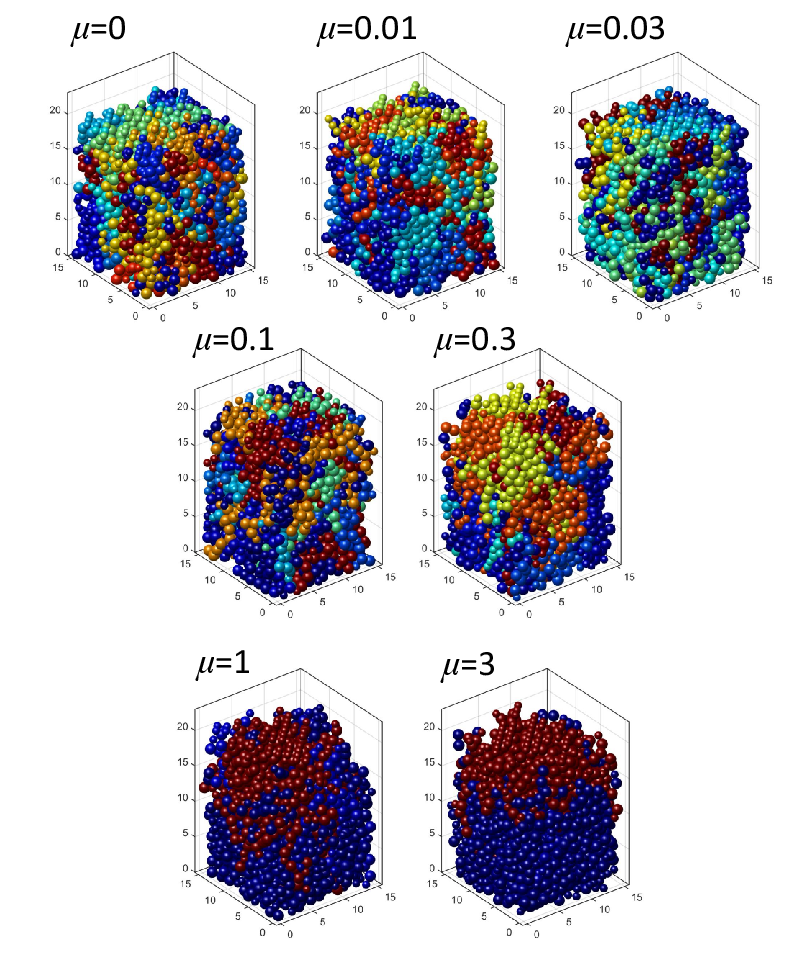}
\caption{Community assignments at all 7 values of interparticle friction $\mu$ and $P=10^{-4}$. Color indicates the network force $\sigma_c$ of each community.  The color scale for each packing ranges from zero (deep blue) to the its maximum value of  $\sigma_{c}$ (red). }
\label{fig:confri}
\end{figure}

To illustrate the methods, we first examine the set of 20 configurations with $P = 10^{-4}$. In Fig.~\ref{fig:confri}, sample community assignments are shown for all seven $\mu$ values. At low values of $\mu$ (top row), small communities dominate, while at large values of $\mu$, there is typically a single large community near the top and many smaller and weaker communities at the bottom. (The multiple low-$\sigma_c$ communities all have similar values of $\sigma_c$ and thereby appear to be the same community (by color) although they are not.) This observation is similar to prior work on the effect of friction coefficient $\mu$ on jamming properties of packings \cite{Silbert2010} in which the bulk packing fraction and coordination number gradually decrease as $\mu$ increases from 0 and they saturate when $\mu$ is larger than 1. This saturation is also reflected in the cumulative distribution figures we examine below.

\begin{figure}
\centerline{\includegraphics[width=0.65\linewidth]{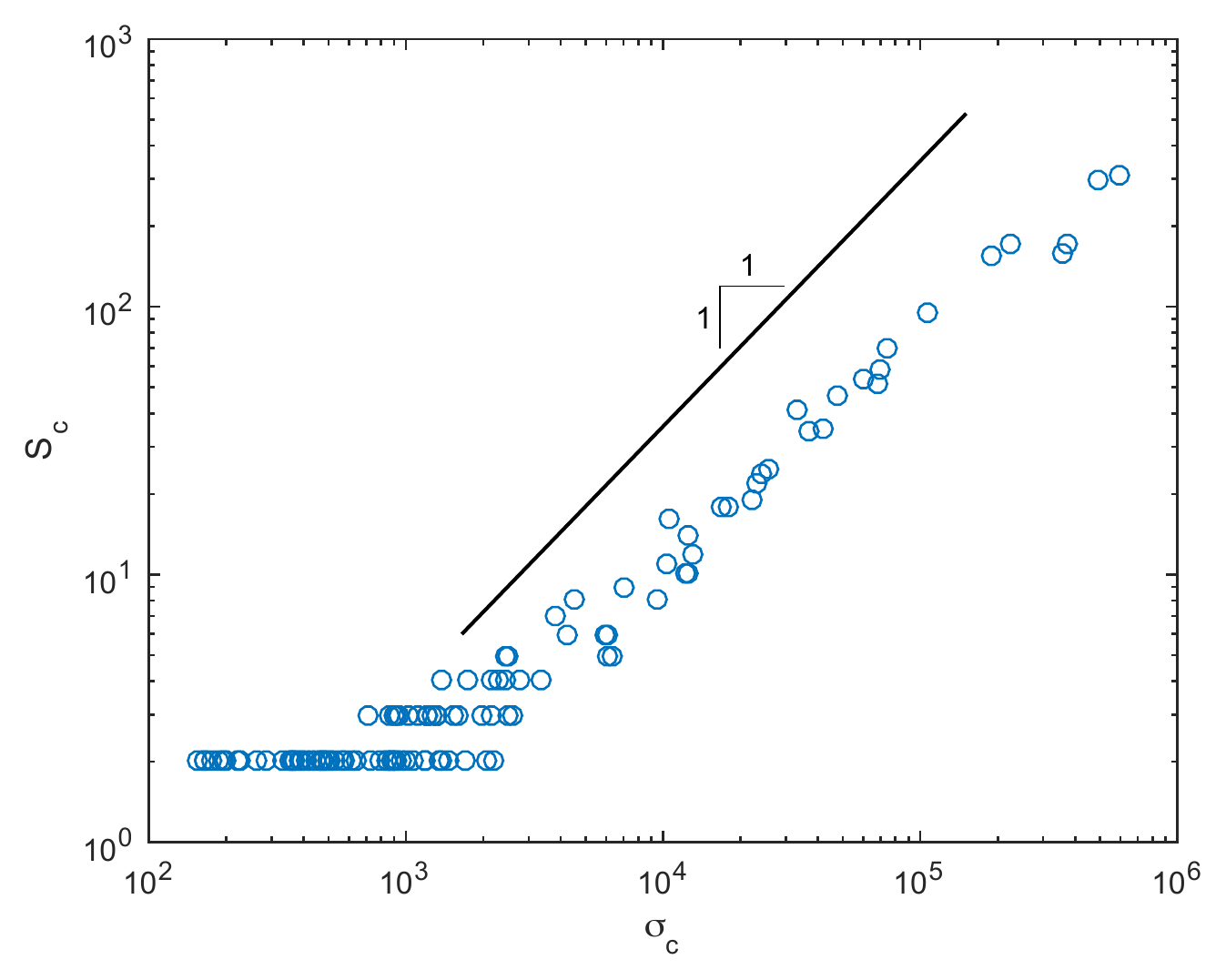}}
\caption{Scatter plot of community size $S_c$ and network force  $\sigma_c$, with each point representing a single community. Data is from a single simulation with $\mu=0.3$ and  $P=10^{-4}$. }
\label{fig:nfsize}
\end{figure}

As shown in Fig.~\ref{fig:nfsize}, we observe that $S_c$ and $\sigma_c$ obey an approximately linear relationship. We examine their relationship under all $\mu$ and $P$ settings and find the same approximate relationship. (Since the mean pressure was already normalized in writing the weighted adjacency matrix $\bf W$, we do not expect a trend in the magnitude of $\sigma_c$.)

\subsubsection{Friction-Dependence} 

\begin{figure}
\centerline{\includegraphics[width=\linewidth]{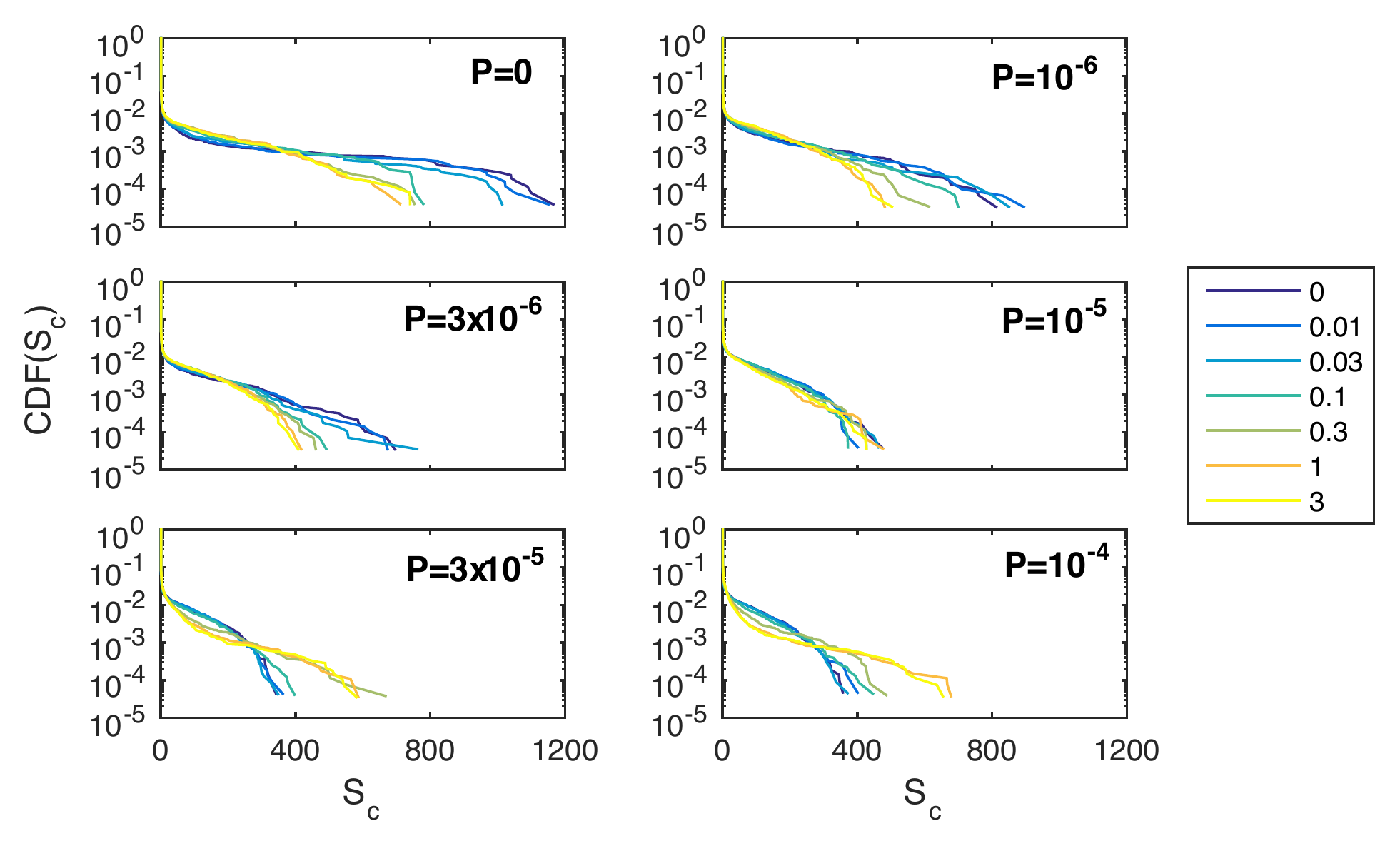}}
\caption{Comparison of the cumulative distribution of community size $S_c$ as a function of $\mu$, where each plot represents the average over 20 simulations at each $P$.}
\label{fig:sizep}
\end{figure}
\begin{figure}
\centerline{\includegraphics[width=\linewidth]{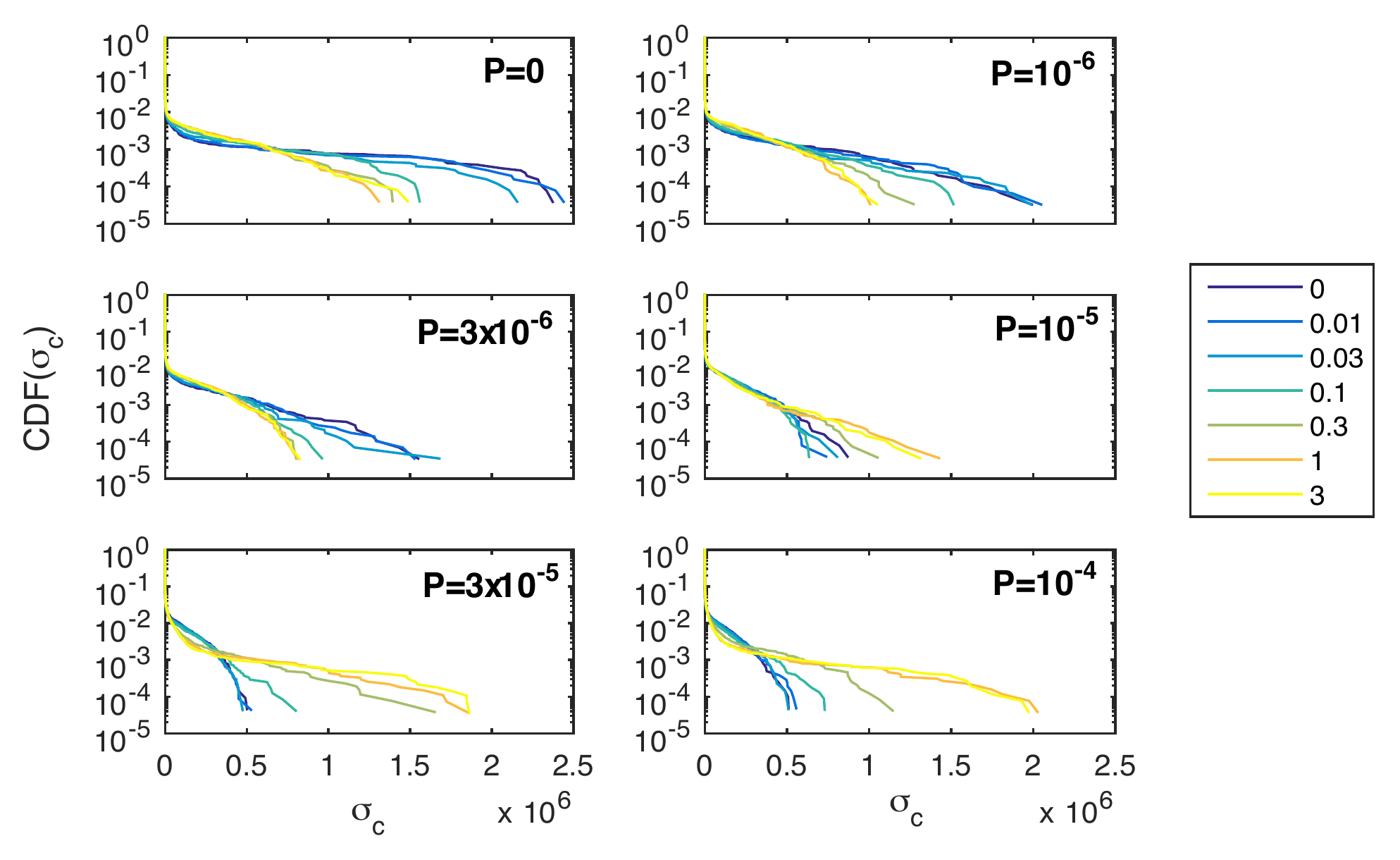}}
\caption{Comparison of the cumulative distribution of network force $\sigma_c$ as a function of $\mu$, where each plot represents the average over 20 simulations at each $P$.}
\label{fig:nfp}
\end{figure}

To understand the friction-dependence, we consider the cumulative distribution function (CDF) of both $S_c$ and network force $\sigma_c$ as a function of $\mu$ at fixed $P$. As shown in Figs.~\ref{fig:sizep} and \ref{fig:nfp}, both quantities show similar behavior, as expected given the strong correlation show in Fig.~\ref{fig:nfsize}. 
For large $\mu$, we observe an approximately exponential distribution. Remarkably, the steepness of the distribution as a function of $\mu$ has  opposite trends at low and high pressure: 
For $P \lesssim 10^5$, the CDF steepens as $\mu$ increases (fewer large/strong communities), while for 
$P \gtrsim 10^5$,  the CDF instead steepens as $\mu$ decreases. 
Thus, $P^* \approx 10^5$ represents a crossover value between two distinct behaviors. Below, we will explore how the heterogeneity of forces (shown illustratively in Fig.~\ref{fig:confri}) causes this effect.

\subsubsection{Pressure-Dependence} 

\begin{figure}
\centerline{\includegraphics[width=\linewidth]{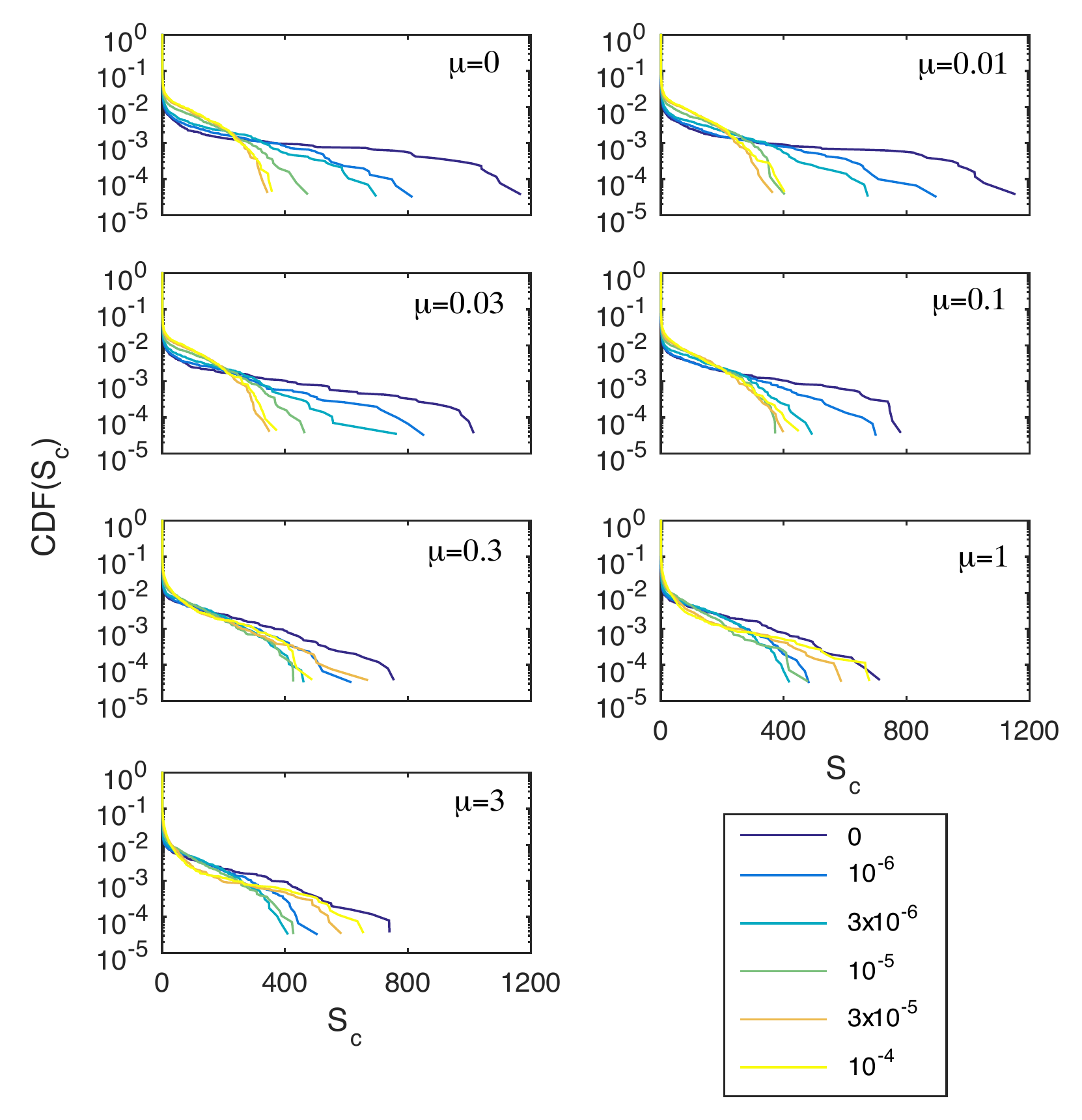}}
\caption{Comparison of the cumulative distribution of community size $S_c$ as a function of $P$, where each plot represents the average over 20 simulations at each $\mu$.}
\label{fig:sizemu}
\end{figure}

\begin{figure}
\centerline{\includegraphics[width=\linewidth]{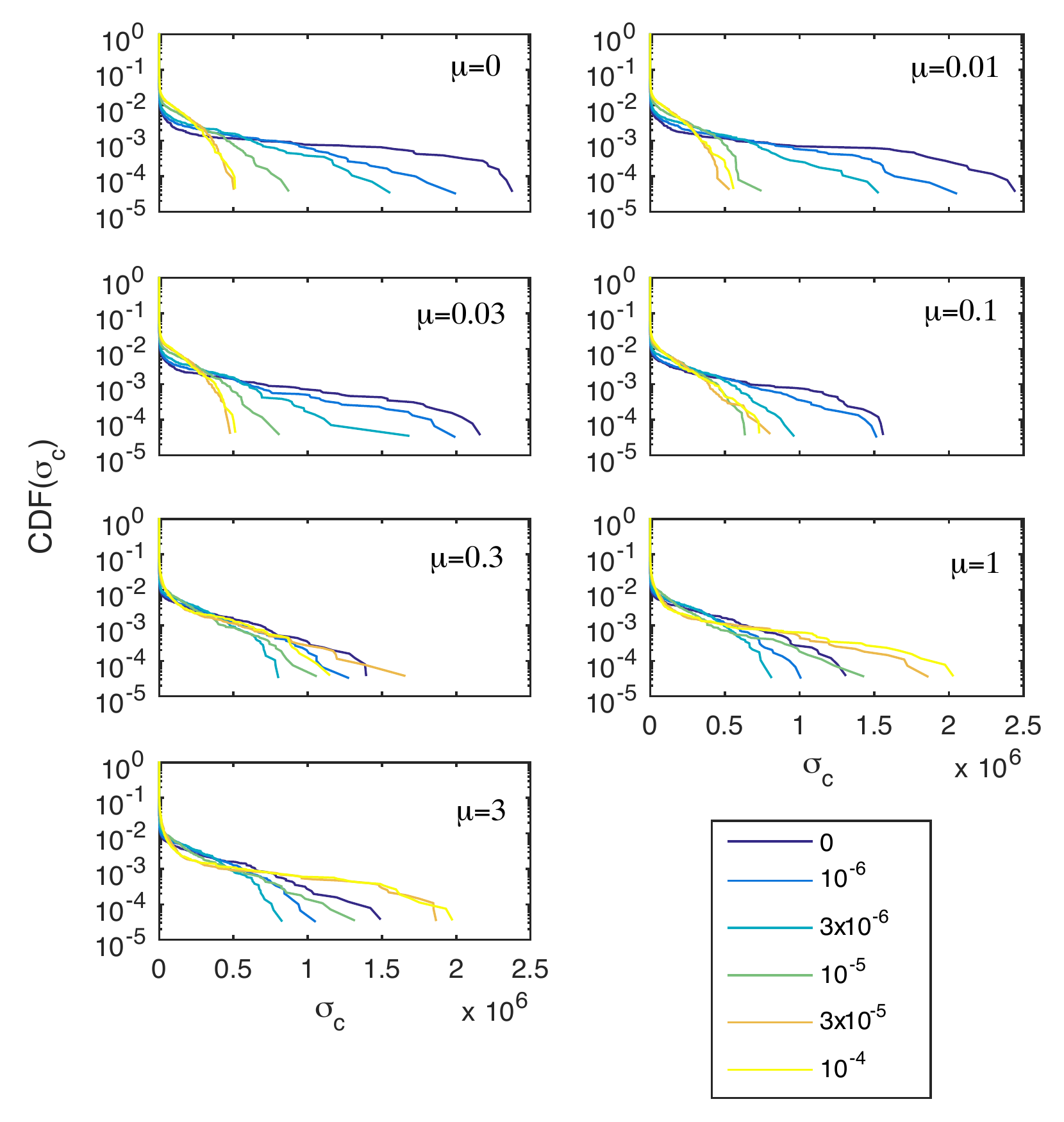}}
\caption{Comparison of the cumulative distribution of network force $\sigma_c$ as a function of $P$, where each plot represents the average over 20 simulations at each $\mu$.}
\label{fig:nfmu}
\end{figure}

Fig.~\ref{fig:sizemu} and \ref{fig:nfmu} show the same CDF data, rearranged to highlight $P$-dependence at fixed $\mu$. 
This configuration highlights the existence of a low-friction regime distinct from the frictional regime, with a transition near $\mu^* \approx 0.1$. 
For $\mu < \mu^*$, the CDFs become much steeper as $P$ is increased. This indicates that the system's forces are becoming more homogeneous at high pressure, as expected \cite{Makse2000,Zhang2005}. 
In contrast, simulations performed at $\mu > \mu^*$ (the frictional regime) show only weak pressure-dependence, with the large-$\sigma_c$ tails fluctuating. 
This may be due either to insufficient statistics, or to changes in the heterogeneity of the system, to be discussed in the next section.

These CDFs of $S_c$ and $\sigma_c$ are similar to those observed in a previous  study of force chains in 2D systems \cite{Bassett2015} using a similar community-detection technique. There, the community size distribution was also  exponential, and here we found that the network force was exponential as well. In addition, both studies saw that  communities are more compact at high pressure. 
This pressure-dependence is in contrast to the work of \citet{Navakas2010}, in which it was observed that community size increases as pressure increases. A key distinction between the two studies is the choice of null model: they used the standard Newman-Girvan null model \cite{Newman2004,Newman2004a} rather than a geometric null model \cite{Bassett2015}, resulting  in domain-like  communities.

\subsubsection{Network Homogeneity} 

We have observed that there is a crossover in community size and strength for both pressure ($P^* \approx 10^5$) and friction ($\mu^* \approx 0.1$), and that this effect appears to be connected to the homogeneity of the force network. To examine this in more detail, we consider the vertical gradient in the community size $S_c$ and its relationship to the relative importance of horizontal vs. vertical forces.

\begin{figure}
\centerline{\includegraphics[width=\linewidth]{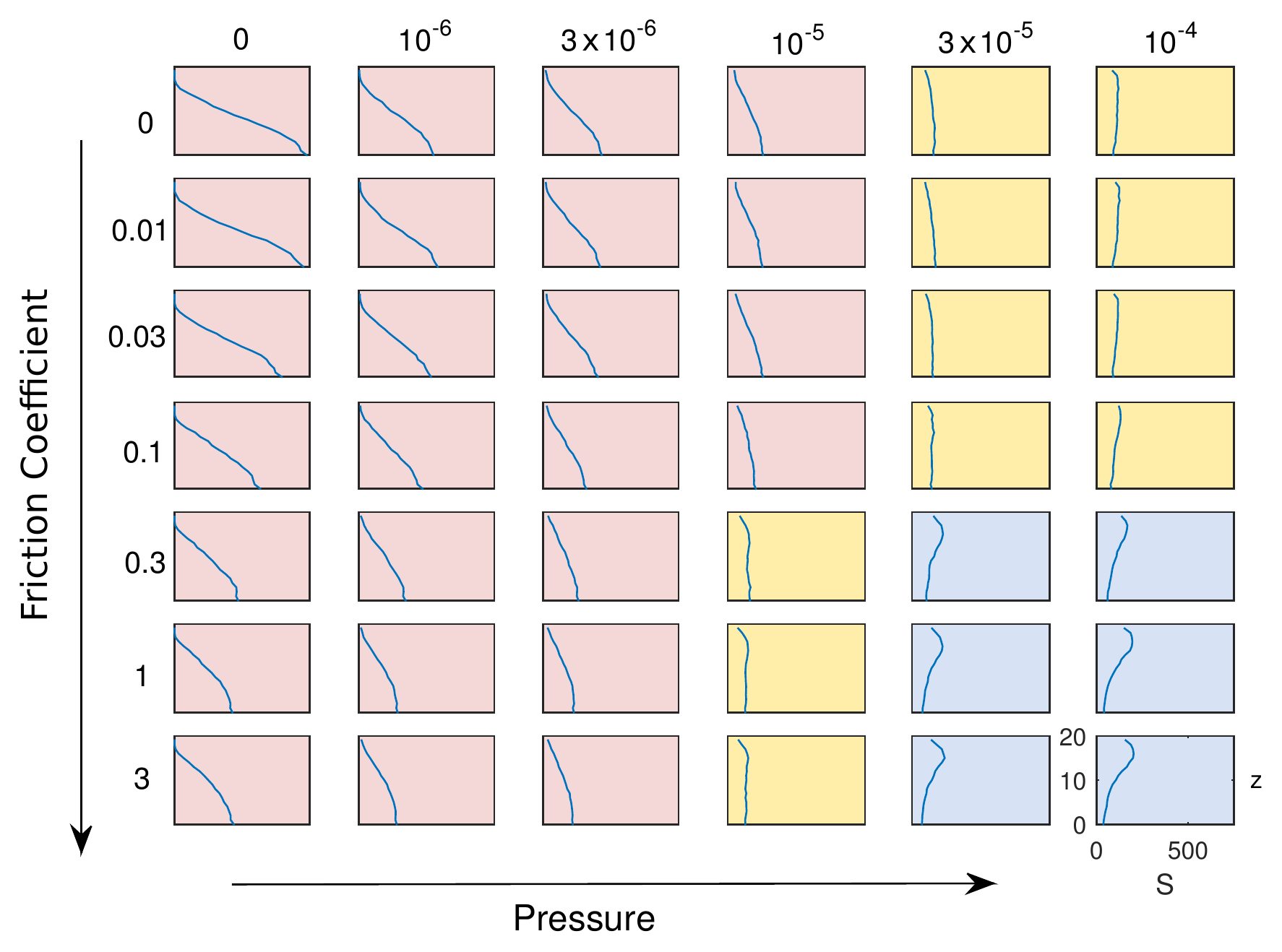}}
\caption{Vertical distribution of average community size $\langle S_c \rangle$ at all values of $\mu$ and $P$ settings, averaged over all 20 simulations. All axis scales are the same. Background colors indicate the slope of the plot: red for negative slope, blue for positive slope, and yellow for uniform community size.}
\label{fig:sizedist}
\end{figure}

Fig.~\ref{fig:sizedist} shows the spatial distribution of average community size $\langle S_c \rangle$ as a function of the vertical position $z$ within the sample, for each pair of $(\mu,P)$ parameters. Averages are calculated on the particle-scale: within  horizontal slice of thickness $d$, we average the $S_c$  of all particles whose centers are within that bin. We observe that the plots fall into three distinct types: negative slope (colored red, largest communities at the bottom), an almost vertical distribution (colored yellow, community size evenly distributed), and positive slope (colored blue, largest communities at the top). As expected from Fig.~\ref{fig:nfsize}, the corresponding plot for $\sigma_c$ is very similar (not shown).

Note that the most homogeneous communities approximately correspond to the $P^*=10^{-5}$ crossover visible in Fig.~\ref{fig:sizep}, suggesting that spatial gradients are important. 
For $\mu > \mu^*$, the largest pressures used were able to reverse the gradient, moving the largest gradients from the bottom to the top of the packing. Note that these two kinds of gradients distinguish the similar-width distributions at $P=0$ and $P=10^{-4}$ in Fig.~\ref{fig:sizemu} when $\mu>\mu^*$. This non-monotonic dependence of heterogeneity on pressure was unexpected. 

\begin{figure}
\centerline{\includegraphics[width=\linewidth]{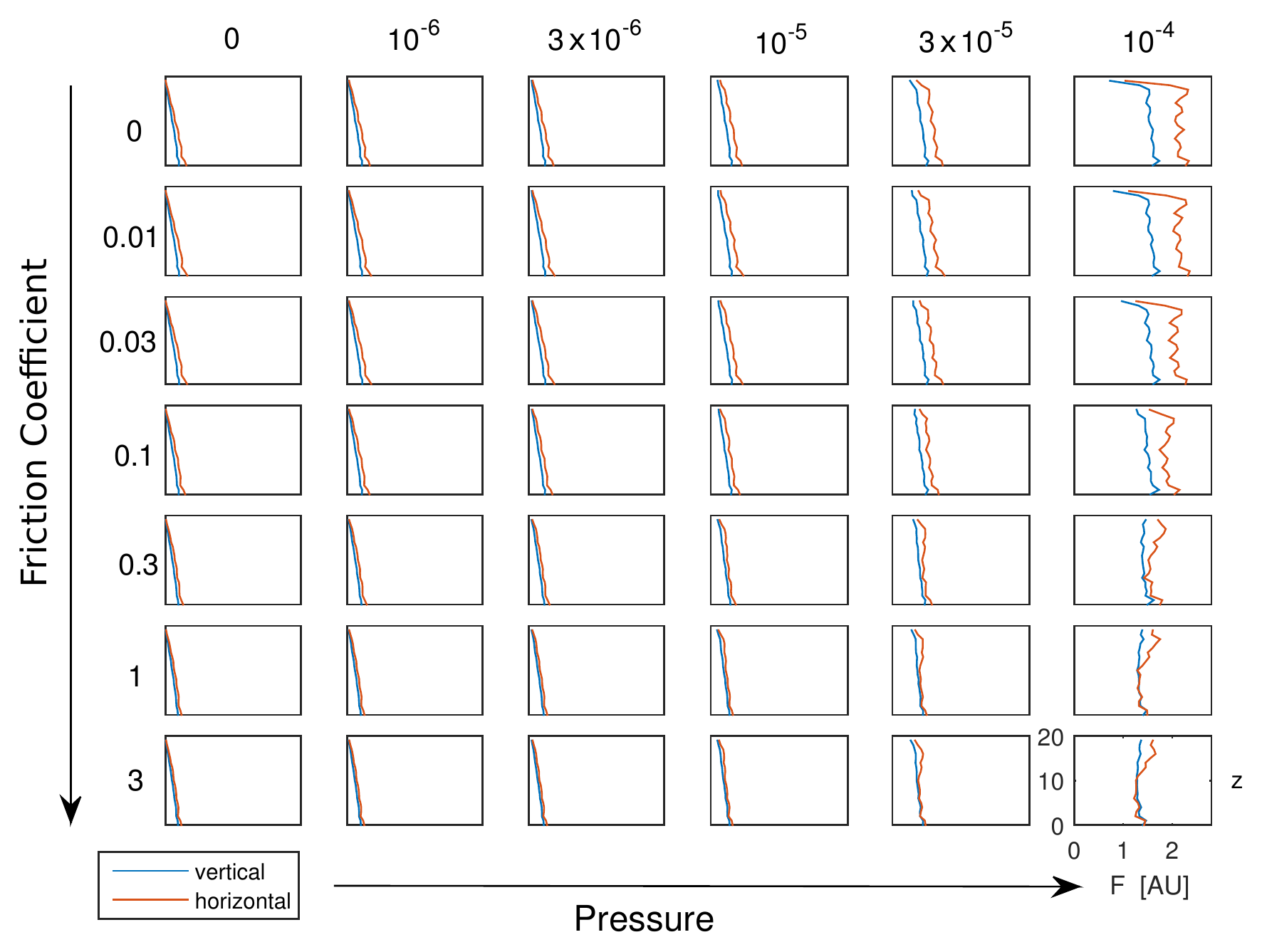}}
\caption{Vertical distribution of average (vertical, horizontal components of the) normal force $\vec f$ at all values of $\mu$ and $P$ settings, averaged over all 20 simulations. All axis scales are the same. 
}
\label{fig:verhor}
\end{figure}

To understand how this effect arises, we consider the relative importance of horizontal and vertical normal interparticle forces as a function of $z$. As done for $\langle S_c \rangle$, we calculate the particle-scale average of the horizontal and vertical components of the vector normal force $\vec f$. As shown in Fig.~\ref{fig:verhor}, the interparticle forces mostly  increase with depth, as would be expected for gravitational loading. (Due to the periodic boundary conditions in the lateral direction, the forces cannot saturate due to the Janssen effect \cite{janssen1895versuche}.) 
This gravitation-loading regime approximately corresponds with the red-shaded plots in Fig.~\ref{fig:sizedist}, and is also visible in the middle column of Fig.~\ref{fig:optgamma}b, where a big, strong community forms at the bottom part of the packing. 
In contrast, for $P > P^*$ the forces are more spatially homogeneous; similar effects have been seen by Makse et al. \cite{Makse2000,Zhang2005}. 
For high $P$ and $\mu$ (the blue-shaded plots), the vertical forces first become  more 
uniform with depth, but eventually develop a force-excess at the top of the packing. This high-force region corresponds to the large communities shown in the bottom row of Fig.~\ref{fig:nfsize}. 
Between these two extreme cases, there is a regime in which the community-size distributions are quite homogeneous (the yellow-shaded plots in  Fig.~\ref{fig:sizedist}). This regime does not precisely correspond with the most homogeneous force distributions shown in Fig.~\ref{fig:verhor}, suggesting that community-detection is sensitive to small changes in the interparticle forces.

\subsection{Community Morphology \label{sec:morph}} 

\begin{figure}
\centerline{\includegraphics[width=0.65\linewidth]{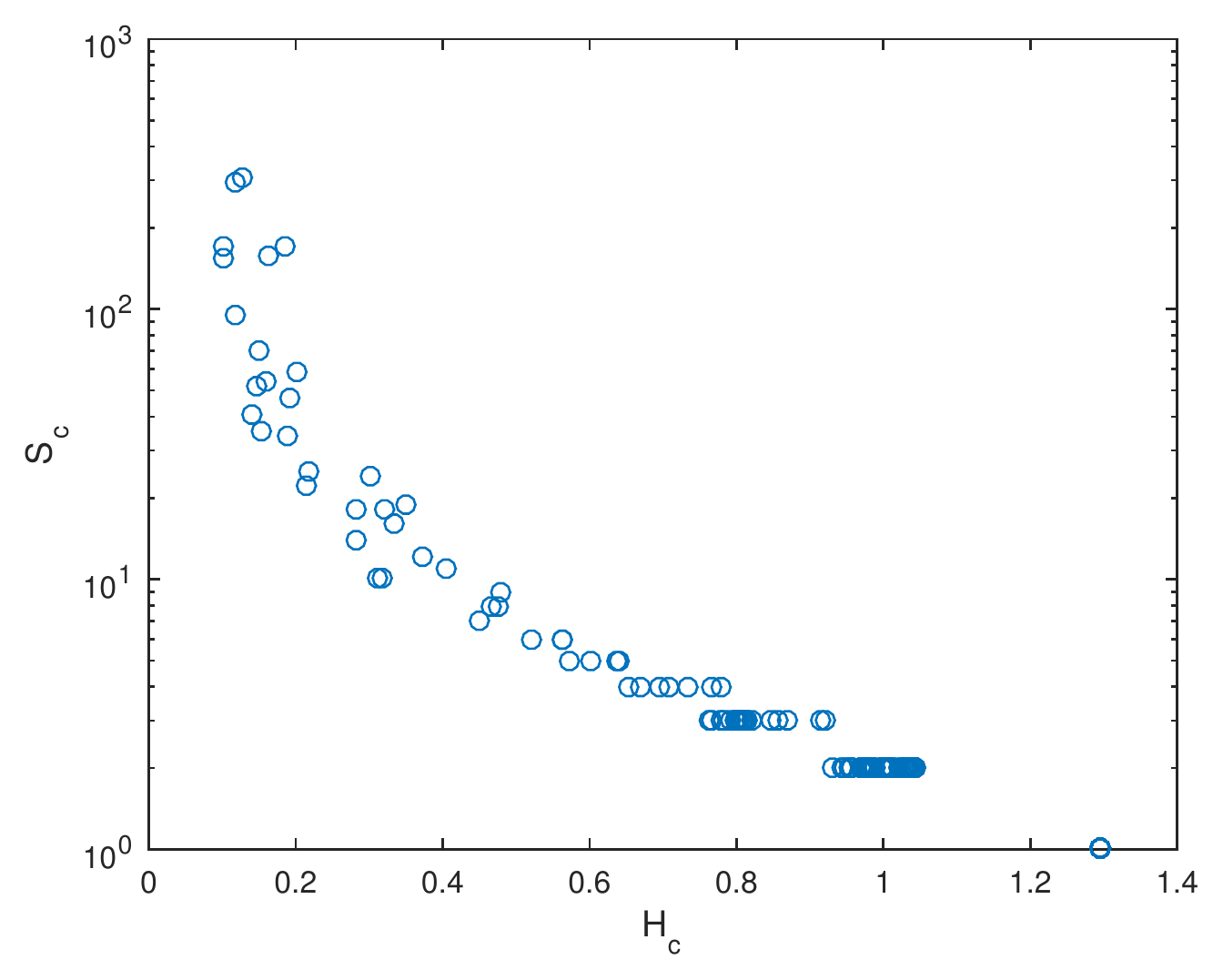}}
\caption{Scatter plot of hull ratio $H_{c}$ as a function of community size $S_c$ for a single simulation at $P=10^{-4}$ and $\mu=0.3$. Values of $H_c > 1$ are possible because we approximate spheres as polyhedra in finding the convex hull.}
\label{fig:hullsize}
\end{figure}

While visual inspection of force chain morphology is possible in 2D systems, it is harder to observe such changes within 3D systems (see Fig.~\ref{fig:lammps}). Therefore, a key benefit of using the  geographical null model (Eq.~\ref{eq:nullmodel})  to detect the communities of particles which form the backbone of the system is to provide a way to quantify changes in the force chain network. The hull ratio $H_c$ (Eq.~\ref{eq:hr}) measures the degree to which the communities are sparse/branched. In this section, we characterize how $H_c$ changes as a function of $\mu$ and $P$. 
As shown in Fig.~\ref{fig:hullsize}, we observe that the largest communities are also the most branched (low $H_c$). An exception to this trend occurs when a large, strong community forms at the top of packing (the blue-shaded plots in Fig.~\ref{fig:sizedist}).

\begin{figure}
\centerline{\includegraphics[width=\linewidth]{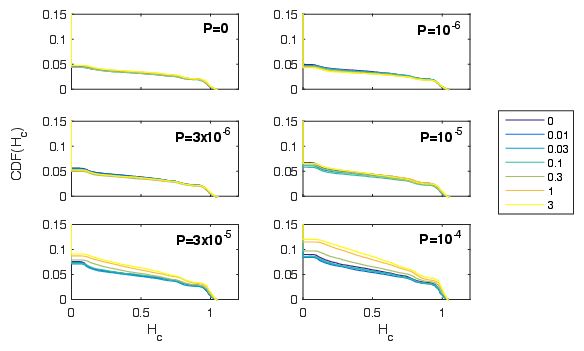}}
\caption{Comparison of the cumulative distribution of hull ratio $H_c$ as a function of $\mu$, where each plot represents the average over 20 simulations at different $P$. Low $H_c$ values correspond to branched and sparse communities, while high $H_c$ values correspond to dense communities.}
\label{fig:hullp}
\end{figure}

Fig.~\ref{fig:hullp} shows the cumulative distributions of hull ratio $H_c$, organized by pressure. Because $\gtrsim 90\%$ of the communities contain only a single particle, these communities are not shown on the plots and have been excluded from calculations of average hull ratio (including in Fig.~\ref{fig:optgamma}). 
For $P \gtrsim P^*$, we observe that the cumulative distributions are sensitive to $\mu$; below $P^*$, they are $\mu$-independent.  In the $\mu$-dependent regime, we see that larger frictional forces contribute to finding more chain-like communities (low $H_c$). Conversely, it is also true that for high $\mu$, the  CDF of $H_c$ is more sensitive to $P$. This is consistent with studies in two dimensions \cite{Bassett2015}, where the community shape for a frictionless packing was less sensitive to pressure than in frictional packings.

\section{Conclusions \label{Conclusions}}

In this paper, we have shown that community detection methods can be successfully applied to 3D granular materials. We define a new quantity, the hull ratio, which characterizes the degree of branching within a community. This quantity allows us to optimize the community detection process by identifying a resolution ($\gamma = 1.1$) where we detect the most-branched features of the system. This resolution is in approximate  agreement with observations in 2D granular systems \cite{Bassett2015}, and is sensible given the normalization of the weighted adjacency matrix $\bf W$.

For packings generated over a range of interparticle friction $\mu$  and pressure $P$, we characterize the detected communities in terms their size, strength, and hull ratio. The first two are found to be largely redundant, and all three depend on $\mu$ and $P$. We find that, as in 2D systems \cite{Bassett2015}, the size  and strength exhibit approximately exponential distributions.  Using these measures, we observe that there  is a crossover in community size and strength for both the pressure ($P^* \approx 10^5$) and friction ($\mu^* \approx 0.1$). In addition, this effect appears to be connected to the homogeneity of the force network.

It is our hope that this technique will prove useful for investigating the statistical properties of force chain networks, by identifying the most important communities of particles. While we have not included tangential forces in this study, including will likely be necessary for addressing questions of mechanical stability.

\begin{acknowledgements}
We are grateful for support from the National Science Foundation (DMR-1206808) and the James S. McDonnell Foundation. The simulations were performed at the NC State High Performance Computing Center. We are grateful to Leo Silbert, Danielle Bassett, and Mason Porter for valuable conversations. 
\end{acknowledgements}


\end{document}